\documentclass[11.0pt,a4paper]{article}
\usepackage{epsfig}

\newlength{\dinwidth}
\newlength{\dinmargin}
\def\ytau{y}

\newcommand{\Le}{\left(}
\newcommand{\Ra}{\right)}

\newcommand{\al}{\alpha}
\newcommand{\T}{T}
\newcommand{\beq}{\begin{equation}}
\newcommand{\eeq}{\end{equation}}
\newcommand{\beqar}[1]{\begin{eqnarray}\label{#1}}
\newcommand{\eeqar}{\end{eqnarray}}
\begin{document}

\title {{~}\\
{\Large \bf  Gluon density and $F_{2}$ functions from BK equation
with impact parameter dependence}\\}
\author{
{\large
S.~Bondarenko$\,{}^{a)}\,$\thanks{Email: sergey@fpaxp1.usc.es}}
 \\[10mm]
{\it\normalsize $^{a)}$ University Santiago De Compostela, Spain }\\}

\maketitle
\thispagestyle{empty}

\begin{abstract}
In this note we fix the preliminary results obtained
in the study of gluon density function of the paper \cite{bom3}.
The LO BK equation for unintegrated gluon density
with impact parameter dependence is considered
in order to fix the parameters of the proposed model.
In particular the form of initial condition for the equations
of proton-proton scattering from \cite{bom3}
is determined, which is similar to the form
of fenomenological GBW ansatz.
The gluon density function and $F_{2}$ function
are also calculated and compared with the results
for the gluon density and $F_{2}$ functions from the
GRV parameterization for different values of $Q^2$.
It is shown, that the results for $F_2$ structure function
of the considered model
are in the good accordance with the results obtained from the
GRV parameterization of parton densities.  
\end{abstract}

\section{Introduction}

 The attempts to understand the
aspects of high energy scattering of nuclei and hadrons 
in terms of QCD BFKL pomerons, \cite{bfkl},
led in the last time to a number of papers
concerning the phenomenological applications of the 
high energy scattering as well as the  pure theoretical 
properties of the theory, \cite{braun1,braun2,bom4,bom,pryl,
bom1,Bond,bom2}.
In present paper we fix the first result obtained in the 
framework of the model proposed in \cite{braun1,braun2,bom}, 
introducing the impact
parameter dependence into the initial conditions and solving the
equations of the model for each point in impact parameter space. 
The found parameters of the model, tuned with the help of DIS, 
are the first step toward the solution of the problem of 
proton-proton scattering considered in \cite{bom3}.

 The DIS process is well described in the frameworks related with the 
BK equation, \cite{bal,kov}, fenomenological models with the
saturation properties and CGC models,
see \cite{GolecBiernat1,GolecBiernat2,GolecBiernat3,Mueller1,
Munier1,Lev1,Lev2,Lev3,Bartels1,Kowalski1,Iancu1,Kowalski2}.
Nevertheless, in rapidity evolution
the impact parameter dependence of the amplitude was
treated only approximately, see for example \cite{Lev1,Lev2}, whereas the fenomenological
models, such as  GBW model \cite{GolecBiernat1,GolecBiernat2},
neglect the evolution of the amplitude with rapidity.
In papers \cite{GolecBiernat3,Kowalski1,Kowalski2} the approximate treatment of
impact parameter dependence of gluon structure function was accounted together with
DGLAP evolution of the function, and in papers \cite{Kut1,Kut2,Kut3}
the factorized from momentum function impact parameter dependence
in modified BK equation was considered. 
But still, the treatment of
impact parameter dependence of the simple BK equation
compatable with the fenomenological models was missed.

  In present calculations of solution of BK equation we include
the impact parameter dependence of the amplitude at
initial values of rapidity and find the
amplitude in each point of impact parameter space,
solving the evolution BK equation with the help of the methods
developed in \cite{bom}, see also the papers \cite{Kut1,Kut2,Kut3}
for similar technics of calculations.
In order to simplify the calculations, the solution is obtained in LO
approximation and we discuss a possible generalization
of the solution till NLO order in the conclusion.
Another important question, which we tried to answer on,  it is a problem
of the form of initial condition function for the BK equation
with impact parameter
dependence. The form of this function is general in the given framework
of interacting BFKL pomerons
and, as we mentioned above, the same function could be used in the proton-proton scattering, see \cite{bom,bom3}.

 Certainly, the   results of calculations
must be clarified with the help of well established results for gluon density
and/or with the help of DIS data. We perform the check of our calculations 
comparing calculated gluon density function (integrated gluon density) and
$F_{2}$ function with the results given by the LO and NLO GRV parameterizations
for DIS data, \cite{GRV}.
This comparison shows, that in the present framework we achieved the
satisfactory description of DIS data.
The model based on  BK equation with impact parameter dependence
shows a good coincidence with GRV parameterization and could be used
as a independent parameterization of unintegrated gluon density, 
\cite{bom3}.
 
 The paper is organizes as follows. In the next section we shortly describe
a formalism of calculations. In Sec.3 and Sec.4 we present the results of calculations
for  $F_2$ structure function and
gluon density function correspondingly.
Section 5 is a conclusion of the paper.

\section{The low-x structure function in the momentum representation }

 In this section we shortly write the main formulae used in our calculations.
The $F_{2}$ structure function of DIS with impact parameter dependence
we define as follows
\beq\label{F1}
F_{2}(x,\,Q^2)\,=\,\frac{Q^2}{4\,\pi^2\,\al}\,\int\,d^2\,b\,
\int\,\frac{d^2\,k}{k^4}\,\frac{f(x,\,k^2,b)}{4\,\pi}\,
\Le\,\Phi_{\T}(k,m^{2}_{q})\,+\,\Phi_{L}(k,m^{2}_{q})\,\Ra\,
\eeq
The unintegrated gluon density function $f(x,\,k^2,b)$ we find solving
the BK
equation for each point in impact parameter space:
\[
\partial_\ytau f(\ytau,k^2,b) =
{N_c \alpha_s \over \pi}\, k^2\, \int {da^2 \over a^2}
\left[
{f(a^2,b)-f(k^2,b) \over |a^2-k^2|} +
{f(k^2,b)\over [4a^4+k^4]^{{1\over 2}}}
\right]
\]
\beq\label{F2}
- {2\pi \alpha_s ^2} \;
\left[
k^2 \int_{k^2} {da^2 \over a^4} \;  f(a^2,b)
\int_{k^2} {dc^2 \over c^4} \;  f(c^2,b)
+ f(k^2,b)\int_{k^2} {da^2 \over a^4} \;
\log\left( {a^2 \over k^2} \right) f(a^2,b)\right]
\eeq
where we introduced the rapidity variable $y=\log(1/x)$.
In Eq.\ref{F2} we assumed that the evolution is local in
the transverse plane, i.e. impact parameter dependence
of $f(\ytau,k^2,b)$ appear only throw the initial condition
for $f(\ytau,k^2,b)$
\beq\label{F3}
f(\ytau\,=\,\ytau_{0}\,,k^2,b)\,=\,f_{in}(k^2,b)\,
\eeq
In order to exclude part of ambiguities in the solution of BK equation
arising due the non included NLO corrections, we perform the following
substitute in the equation
\beq\label{F4}
f(\ytau,k^2,b)\rightarrow\,\frac{f(\ytau,k^2,b)\alpha_s(k^2)}{\alpha_s}=
\frac{\tilde{f}(\ytau,k^2,b)}{\alpha_s}\,,
\eeq
obtaining
\[
\partial_{\tilde{\ytau}} \tilde{f}(\tilde{\ytau},k^2,b) =
{N_c \over \pi}\, k^2\, \int {da^2 \over a^2}
\left[
{\tilde{f}(a^2,b)-\tilde{f}(k^2,b) \over |a^2-k^2|} +
{\tilde{f}(k^2,b)\over [4a^4+k^4]^{{1\over 2}}}
\right]
\]
\beq\label{F5}
- {2\pi} \;
\left[
k^2 \int_{k^2} {da^2 \over a^4} \;  \tilde{f}(a^2,b)
\int_{k^2} {dc^2 \over c^4} \;  \tilde{f}(c^2,b)
+ \tilde{f}(k^2,b)\int_{k^2} {da^2 \over a^4} \;
\log\left( {a^2 \over k^2} \right) \tilde{f}(a^2,b)\right]
\eeq
where $\tilde{y}=\alpha_s\,y$. The  value of $\alpha_s$ is a constant
in the LO approximation and
we consider $\alpha_s$ as the parameter of the model which
must be determined from the fit of  DIS data.

 The impact factors in Eq.\ref{F1} are usual impact factors of the problem with
three light quarks flavors of equal mass included.
They are the following (see \cite{Kwi1} for example):
\beq\label{F6}
\Phi_{L}(k,m_{q}^{2})\,=\,32\,\pi\,\al\,
\sum^{3}_{q=1}\,e^{2}_{q}\,
\int_{0}^{1}\,d\rho d\eta \,\,
\frac{k^2\,\eta\,(1-\eta)\,\rho^{2}\,(1-\rho)^{2}\,Q^{2}\,}
{\Le\,Q^{2}\rho(1-\rho)+k^{2}\eta(1-\eta)+m_{q}^{2}\Ra\,
\Le\,Q^{2}\rho(1-\rho)+m_{q}^{2}\Ra\,}
\eeq
and
\[
\Phi_{\T}(k,m_{q}^{2})\,=\,4\,\pi\,\al\,
\sum^{3}_{q=1}\,e^{2}_{q}\,
\int_{0}^{1}\,d\rho d\eta \,\cdot\,
\]
\beq\label{F7}
\,\cdot\,\frac{k^2\,Q^{2}\Le\,\rho^2 +(1-\rho)^{2}\,\Ra\,\rho\,
(1-\rho)\Le\eta^2 +(1-\eta)^{2}\,\Ra\,+
k^2\,\Le \rho^2 +(1-\rho)^{2}\,\Ra\,m_{q}^{2}\,+
4\,\rho\,(1-\rho)\,\eta(1-\eta)\,m_{q}^{2}}{
\Le\,Q^{2}\rho(1-\rho)+k^{2}\eta(1-\eta)+m_{q}^{2} \Ra\,
\Le\,Q^{2}\rho(1-\rho)+m_{q}^{2}\Ra\,}
\eeq
We exclude $\alpha_s$ from the definition of the impact factors
rewriting the $F_{2}$ structure function in the following way
\beq\label{F8}
F_{2}(x,\,Q^2)\,=\,\frac{Q^2}{4\,\pi^2\,\al}\,\int\,d^2\,b\,
\int\,\frac{d^2\,k}{k^4}\,\frac{\tilde{f}(x,\,k^2,b)}{4\,\pi}\,
\Le\,\Phi_{\T}(k,m^{2}_{q})\,+\,\Phi_{L}(k,m^{2}_{q})\,\Ra\,=\,
\,\frac{Q^2}{4\,\pi^2\,\al}\,\int\,d^2\,b\,S(x,\,b,\,Q^2)\,
\eeq
Due the including quark masses in the calculations, the
rapidity variable $y$ (Bjorken x) in BK equation is also modified,
see details in \cite{GolecBiernat1,GolecBiernat2}.
For each fixed rapidity $y$ of the process, the value rapidity taken in BK equation
is changed
\beq\label{F9}
y\,\rightarrow\,y\,-\,\ln (1\,+\,\frac{4\,m^{2}_{q}}{Q^2})
\eeq

\begin{figure}[hptb]
\begin{tabular}{ c c}
\psfig{file=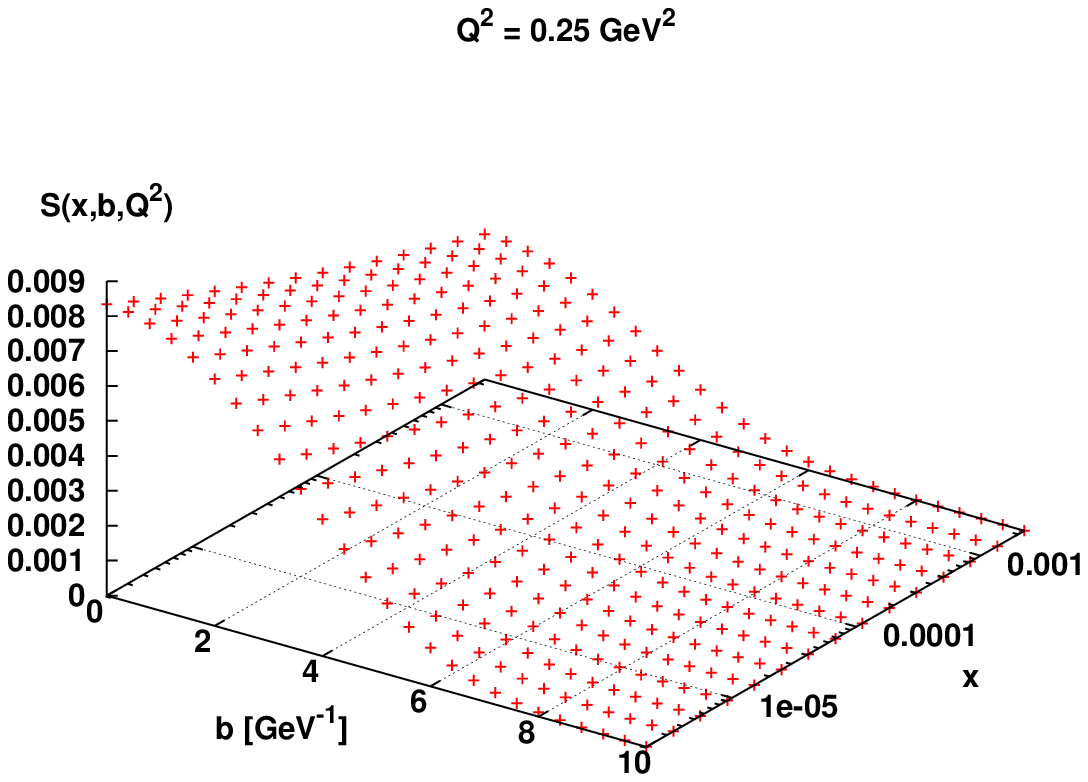,width=80mm} &
\psfig{file=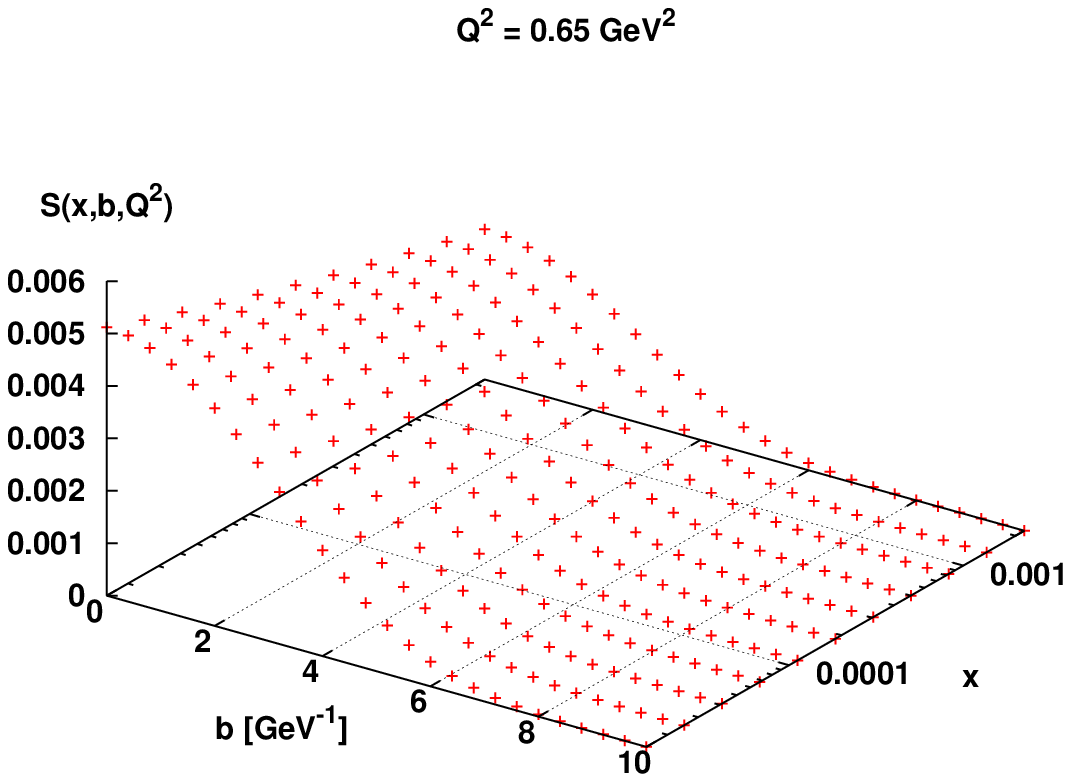,width=80mm}\\
\psfig{file=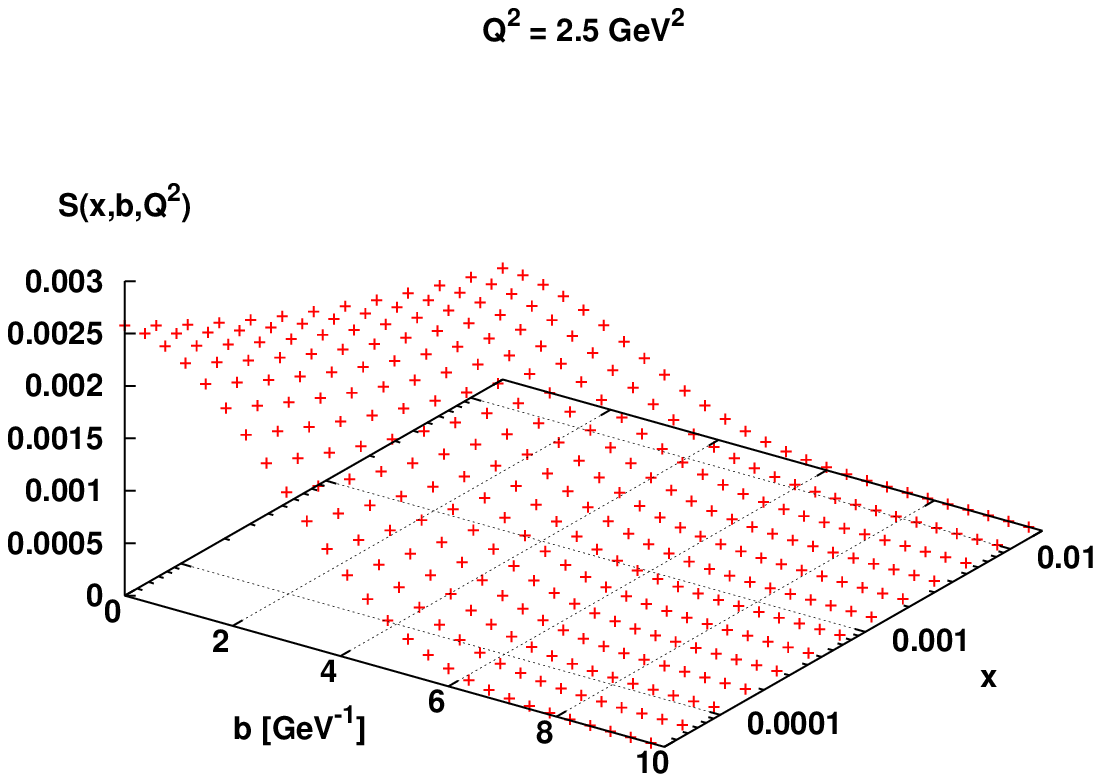,width=80mm} &
\psfig{file=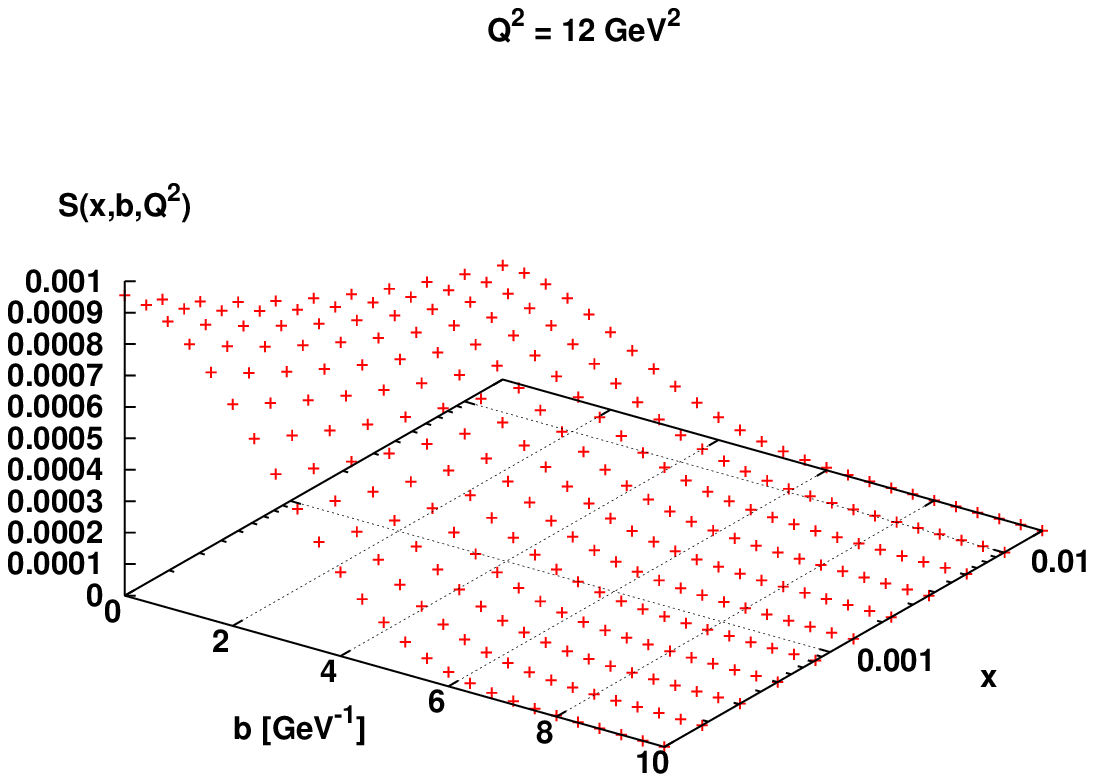,width=80mm}\\
\psfig{file=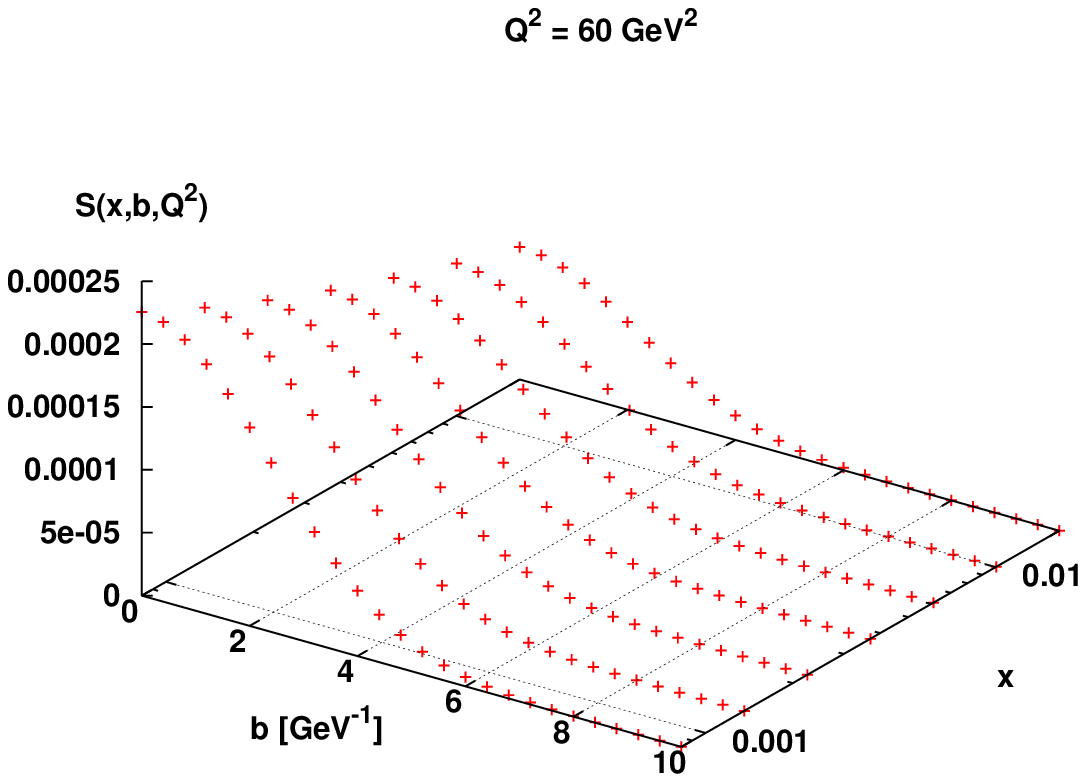,width=80mm} &
\psfig{file=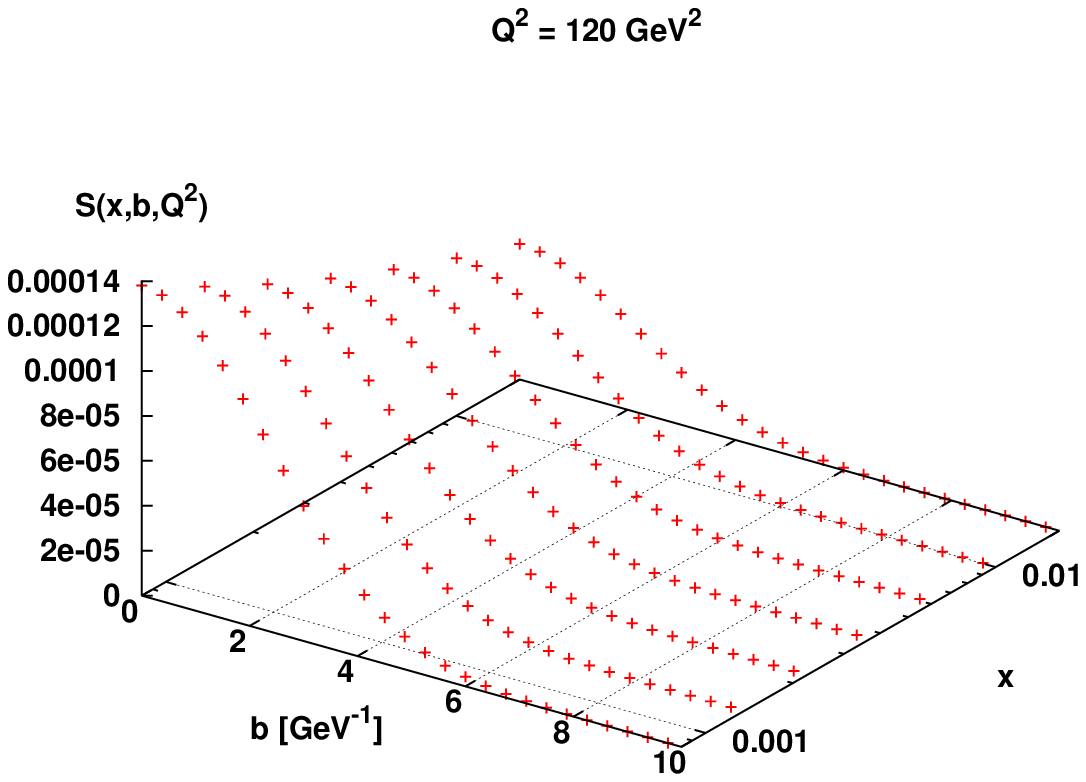,width=80mm}\\
\end{tabular}
\caption{\it The impact parameter profile
of $S(x,\,b,\,Q^2)$ as a function of $x$.}
\label{3dplot}
\end{figure}

 The form of the function $\tilde{f}(\ytau,k^2,b)$ at initial rapidity,
i.e. initial condition for the BK equation Eq.\ref{F5},
has been borrowed from the form of GBW ansatz,
\cite{GolecBiernat1,GolecBiernat2}, with introduced
impact parameter dependence
\beq\label{F10}
\tilde{f}(\ytau\,=\,\ytau_{0},k^2,b)\,=\,\frac{3}{4\pi^2\,}\,
k^4\,R_{0}^{2}\,e^{b^2/R_{p}^{2}}\,
exp(-k^2\,R_{0}^{2}\,e^{b^2/R_{p}^{2}}\,)
\eeq
Additionally to the $\alpha_s$ from the BK equation Eq.\ref{F5}
there are three more parameters, which are
initial rapidity of evolution $\ytau_{0}$, radius of the proton
$R_{p}^{2}$ and "saturation" radius $R_{0}^{2}$. These parameters
must be found from the fitting of DIS data and
they are presented in the next section. The plots of the
functions $S(x,\,b,\,Q^2)$ from Eq.\ref{F8} are given
in Fig~\ref{3dplot}.

\section{$F_{2}$ function and the parameters of the model}

\begin{figure}[hptb]
\begin{tabular}{ c c}
\psfig{file=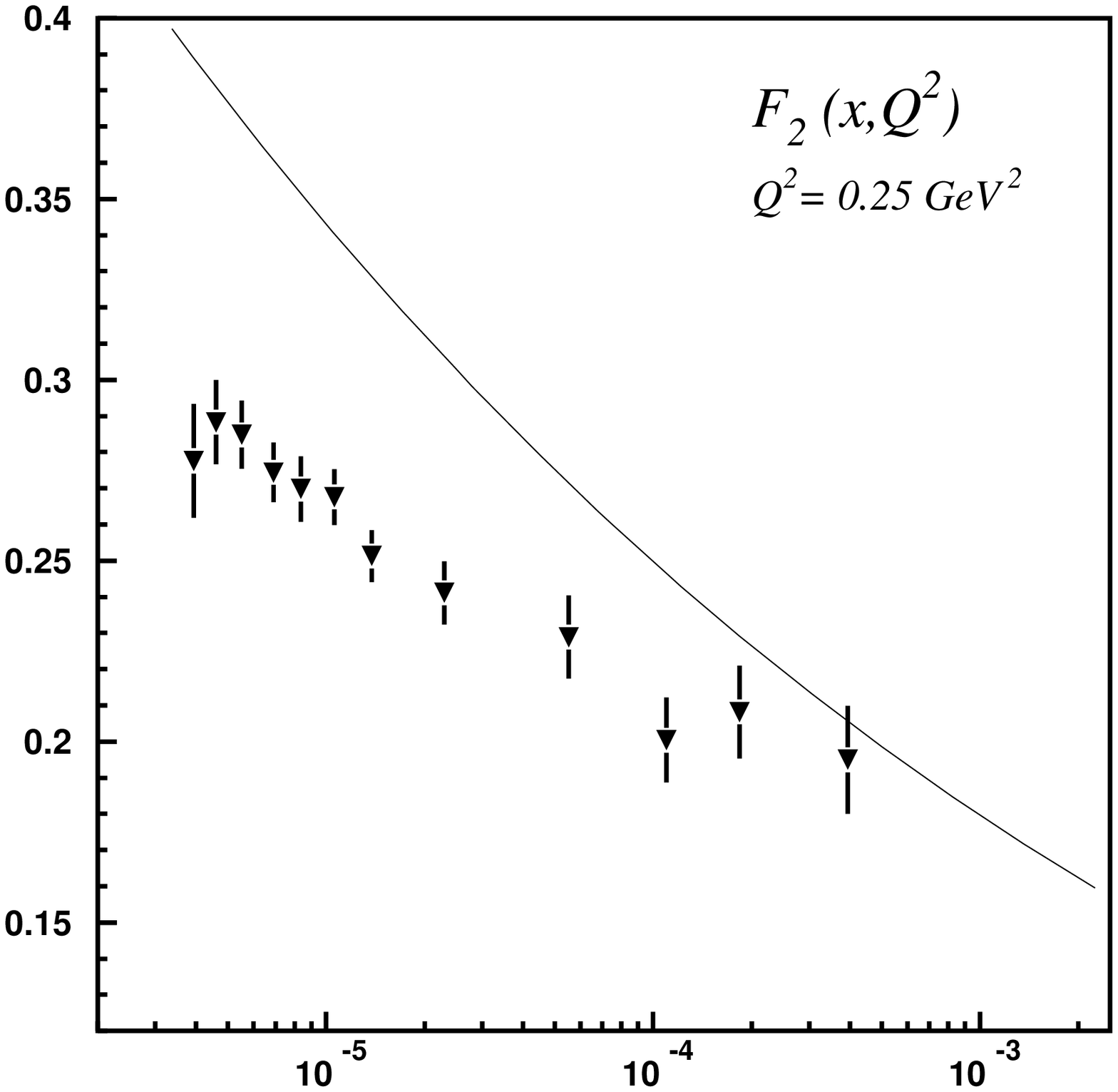,width=75mm} &
\psfig{file=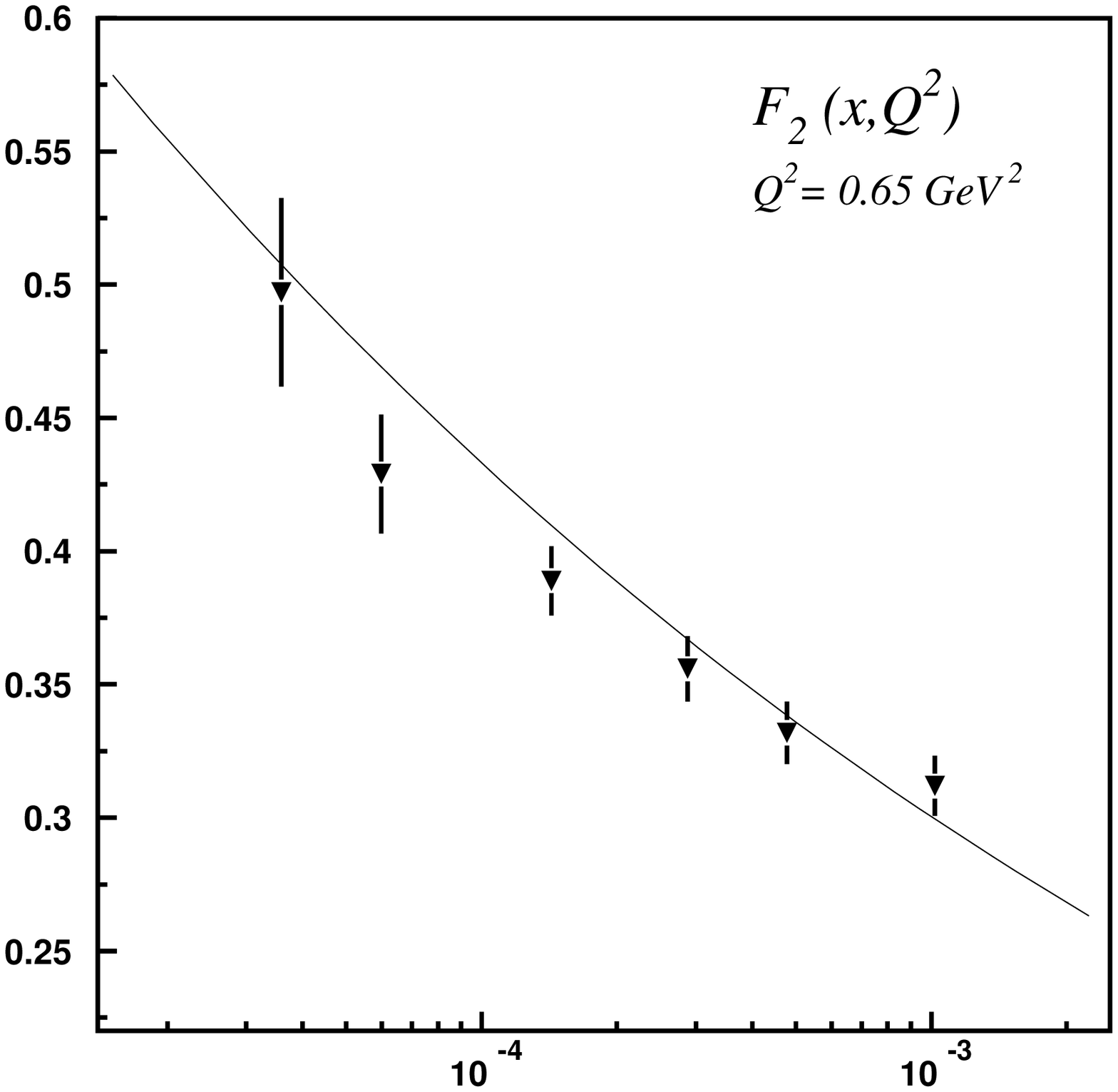,width=75mm}\\
\psfig{file=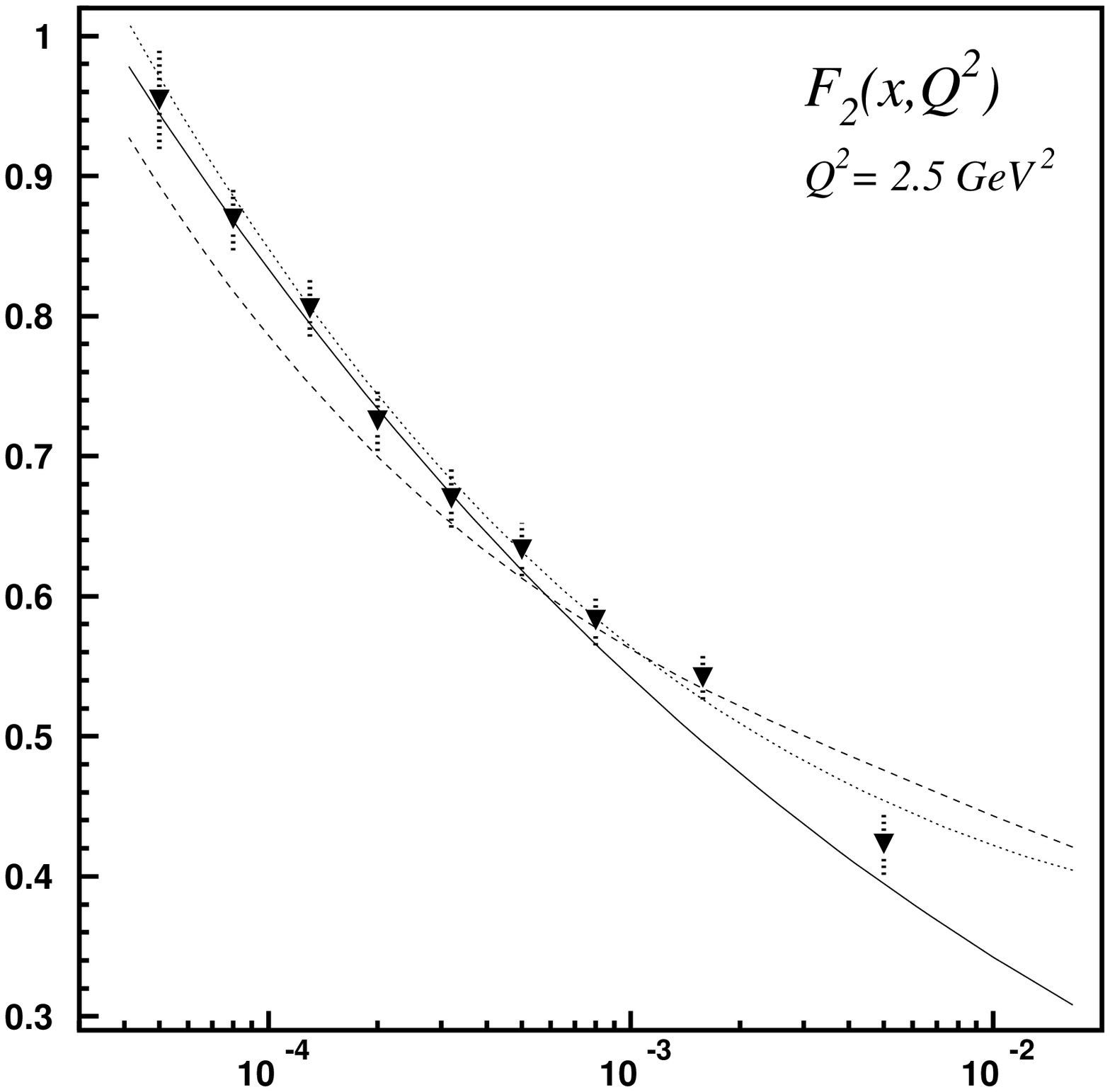,width=75mm} &
\psfig{file=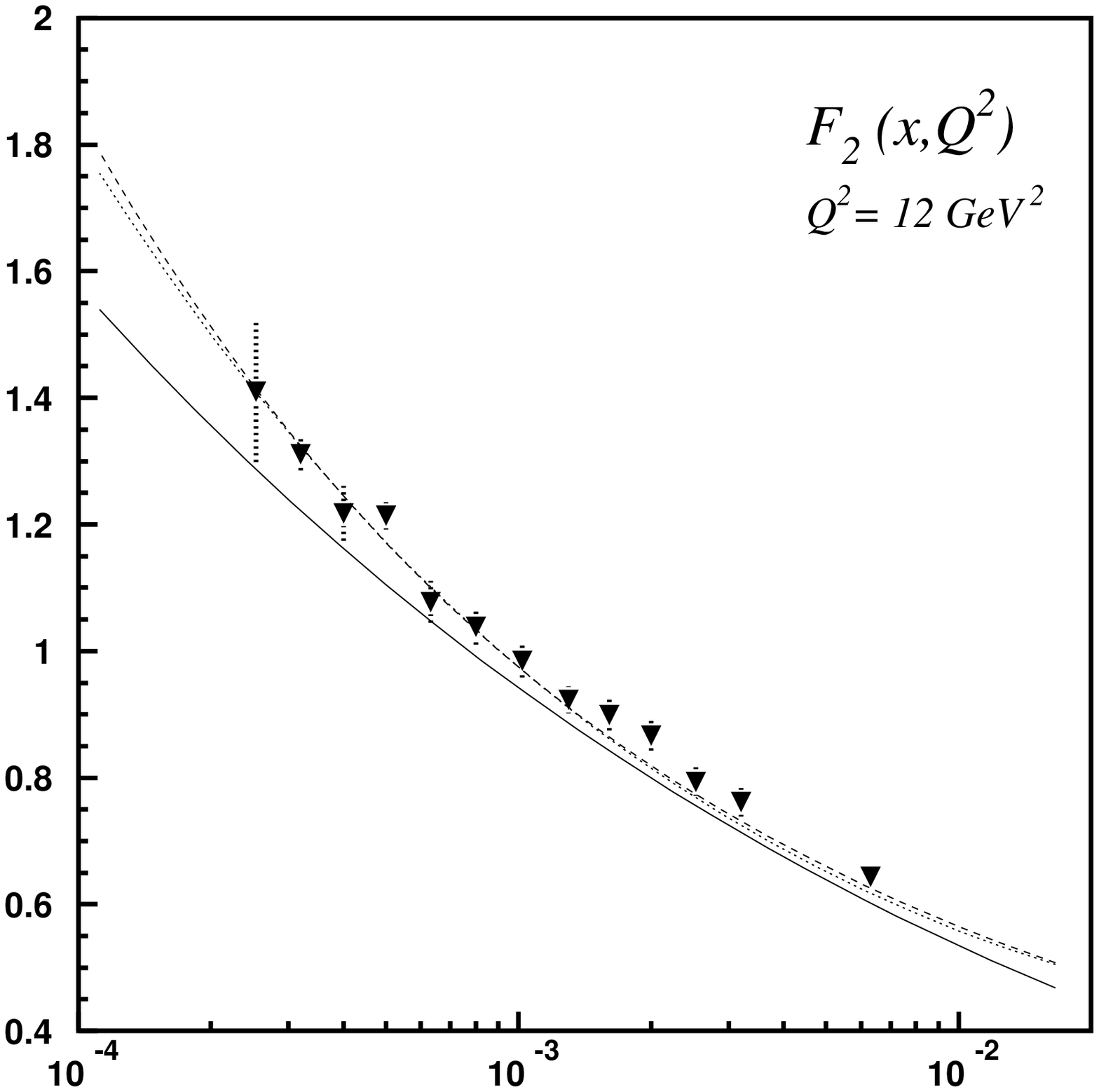,width=75mm}\\
\psfig{file=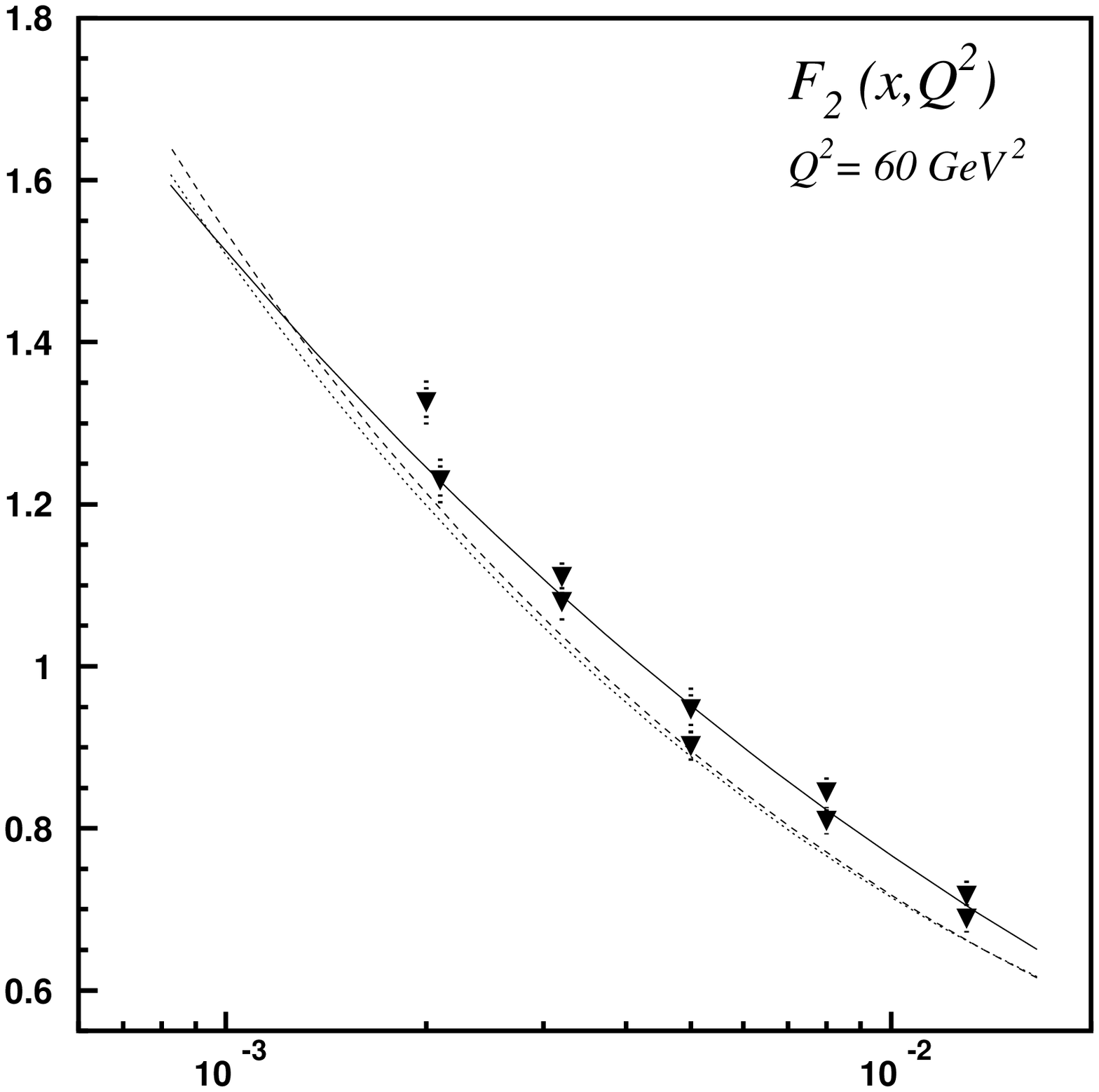,width=75mm} &
\psfig{file=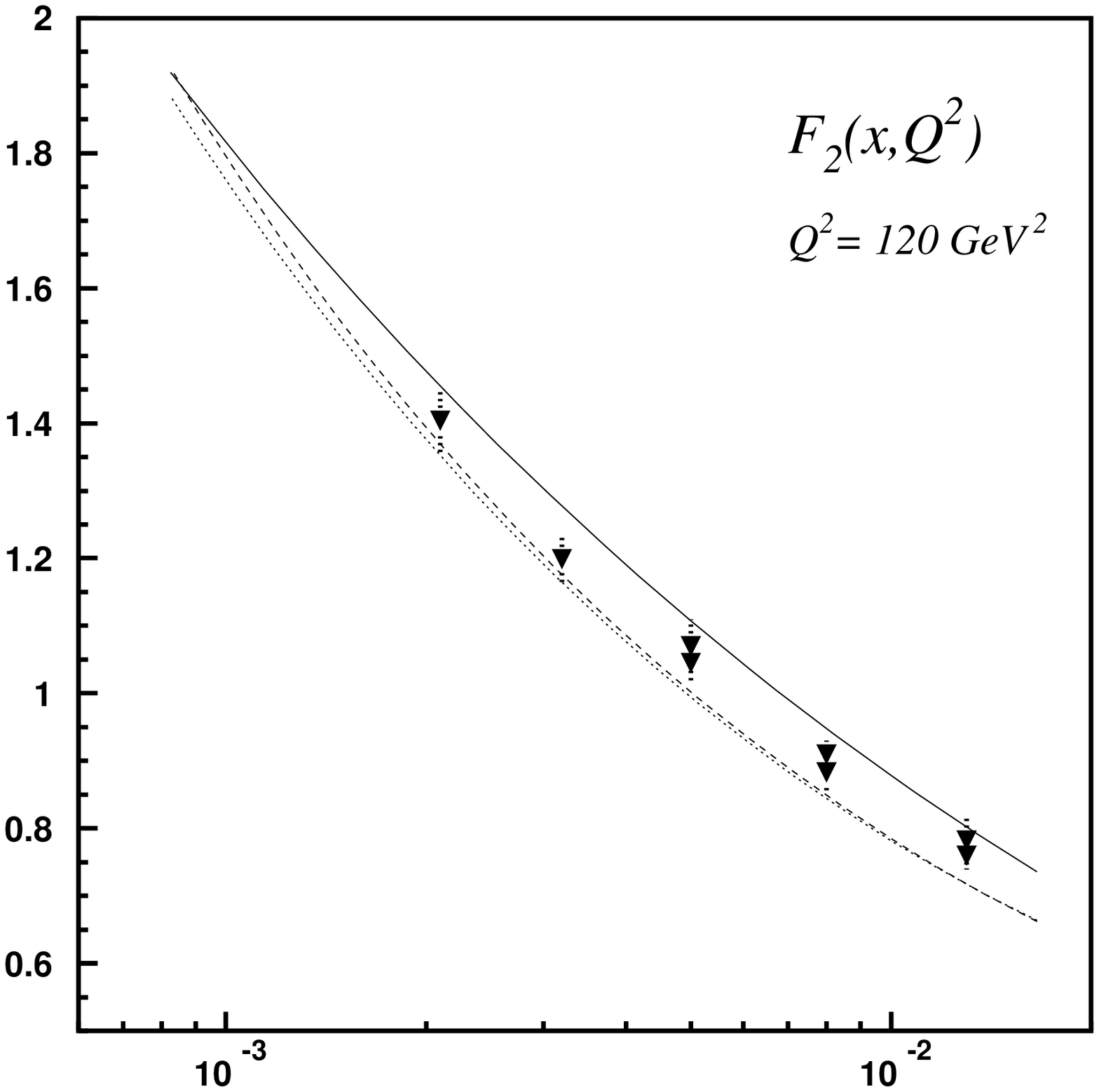,width=75mm}\\
\end{tabular}
\caption{\it The $F_2$ data for different values of $Q^2$:
the present model results (solid lines), LO GRV
parameterization (dashed lines) and NLO GRV (doted lines)
as functions of $x$. The GRV results are restricted by
$Q^2\,>\,0.8\,GeV^2$.}
\label{F2data}
\end{figure}

 The parameters of the model we determine fitting the DIS data
for the total cross section and $F_{2}$
structure function, \cite{DIS}. 
In this  note we present the parameters
of the model and results of the calculations of $F_{2}$
for only few values of $Q^2$ with only three light quarks flavors included,
more results, including the
application of the model to the "soft" proton-proton scattering,
will be presented in the mentioned above paper \cite{bom3}.
The Table~\ref{Param}
shows found values of the parameters of the model
and plots Fig~\ref{F2data} present the results of calculations for the
$F_{2}$ structure function.
\begin{table}[hptb]
\begin{center}
\begin{tabular}{|c|c|c|c|c|}
\hline
\, & \,& \,& \, &\,\\
$y_0\,(x_0)\,$ &  $R^{2}_{p}\,\,(GeV^{2})$ & $R_{0}^{2}\,\,(GeV^{2})$ &
$\alpha_{s}$ & $m_{q}^{2}\,\,(GeV^{2})$ \\
\, & \,& \,& \, &\,\\
\hline
\, & \,& \,& \, &\,\\
$3.1\,\,(0.045) $ & 7.9 & 2.12 & 0.108 & 0.008 \\
\, & \,& \,& \, &\,\\
\hline
\end{tabular}
\caption{\it The parameters of the model.}
\label{Param}
\end{center}
\end{table}
It must be mentioned, that instead the
the $\lambda\,,\,\lambda_{GBW}$ parameters which determines the
energy dependence of the saturation radius in the CGC and GBW models, see 
\cite{GolecBiernat1,GolecBiernat2,Iancu1,Kowalski2}, in our calculations 
the $\alpha_s$ is the parameter which determines the energy behavior of the
gluon density functions. The smallness of obtained value of $\alpha_s$
is explained by the LO precision of the calculations and by the
variable change Eq.\ref{F4}. In present scheme the value of $\alpha_s$
determines the evolution length in rapidity space independently on values
of $Q^2$, i.e. this is some "averaged" value for $\alpha_s$
found from the data fitting.
The found value of quark mass is also different from the
numbers of \cite{GolecBiernat1,GolecBiernat2} for example,
being nevertheless in the range of possible
quark masses of light quarks flavors used in  \cite{Kowalski1,Kowalski2}.
The radius of the proton from Table~\ref{Param}  is close to the experimental
value of the proton shape found from the t-distribution of
$J/\psi$ meson of \cite{DIS}, in fact the results of this  measurements
restrict the possible numericall values of this parameter.

 The main difference
of our calculations from the results of other models are
the results for $F_2$ structure function at very small values of $Q^2$.
The two top plots
of Fig~\ref{F2data}, calculated at small values of
$Q^2$, show that our model fails to describe the $F_2$ data
at $Q^2\,\sim\,0.25,GeV^2 $ . It means, that saturation effects which
provides the description of the low $Q^2$ data in "canonical"
saturation model, such as  \cite{GolecBiernat1} for example ,
in our framework are not so strong as there.
The possible reasons for such a
deviations from the "normal" results obtained at low
$Q^2$ and small $x$ we discuss in
conclusion.
Nevertheless, the comparison of the obtained results with the GRV parameterization results for $F_2$ function for $Q^2\,>\,1\,GeV^2$
shows a good coincidence, they stay
in the limits of differences between the results of GRV parameterization with
the results given by other parton density parameterizations, such as 
\cite{STEQ}for example .

\section{Integrated gluon density function}

\begin{figure}[hptb]
\begin{tabular}{ c c}
\psfig{file= 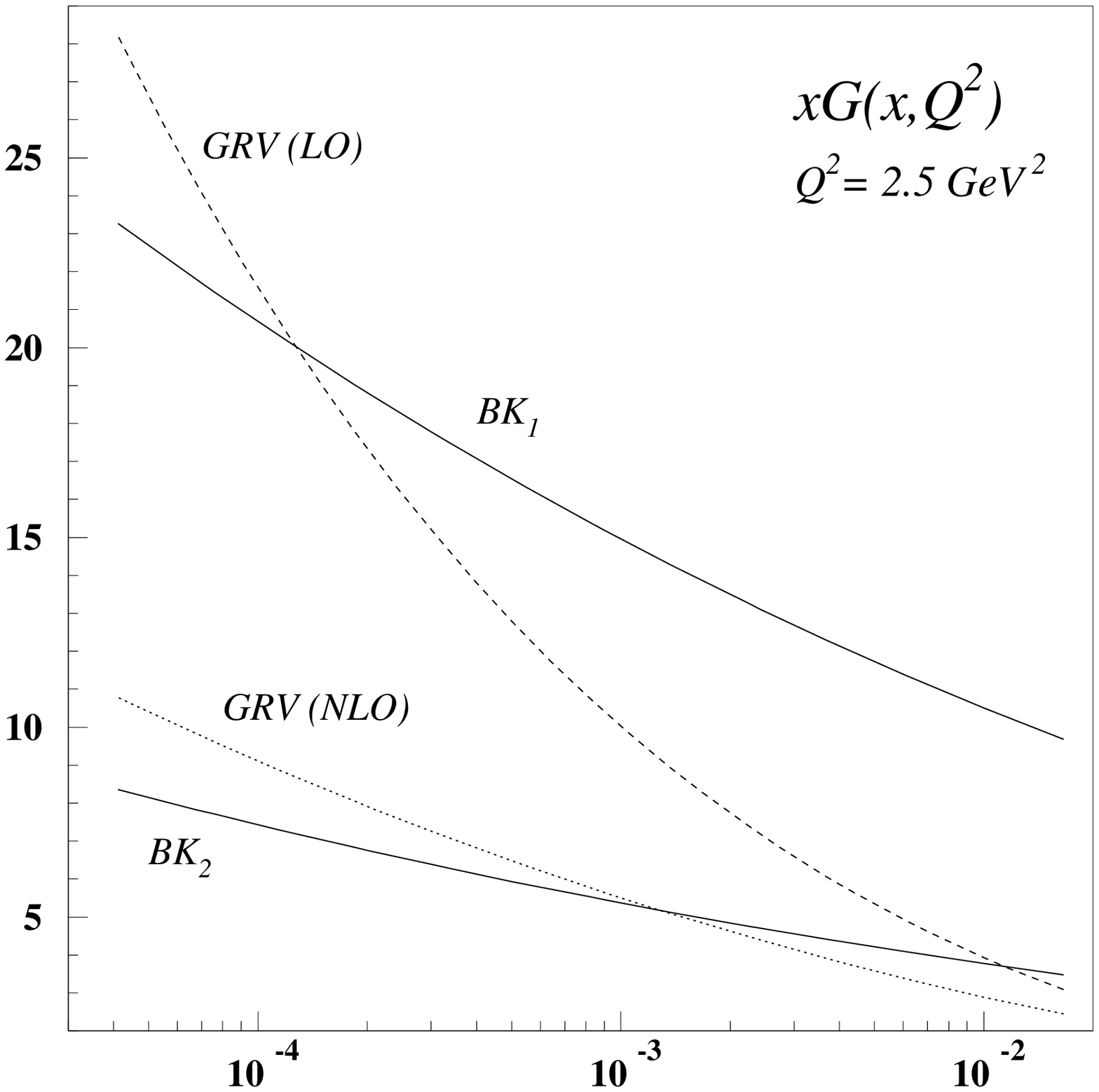,width=85mm} &
\psfig{file= 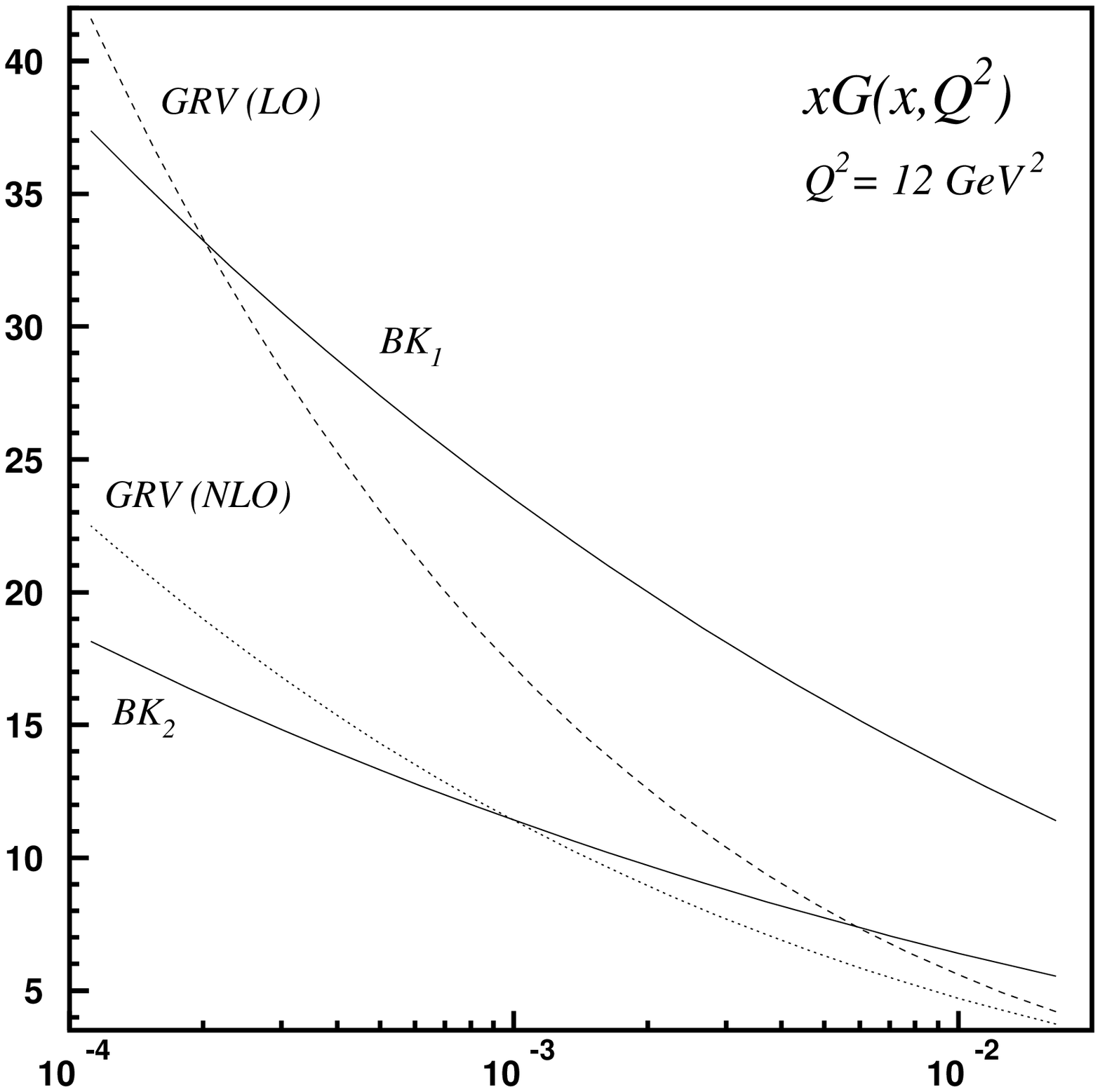,width=85mm}\\
 &  \\
\psfig{file= 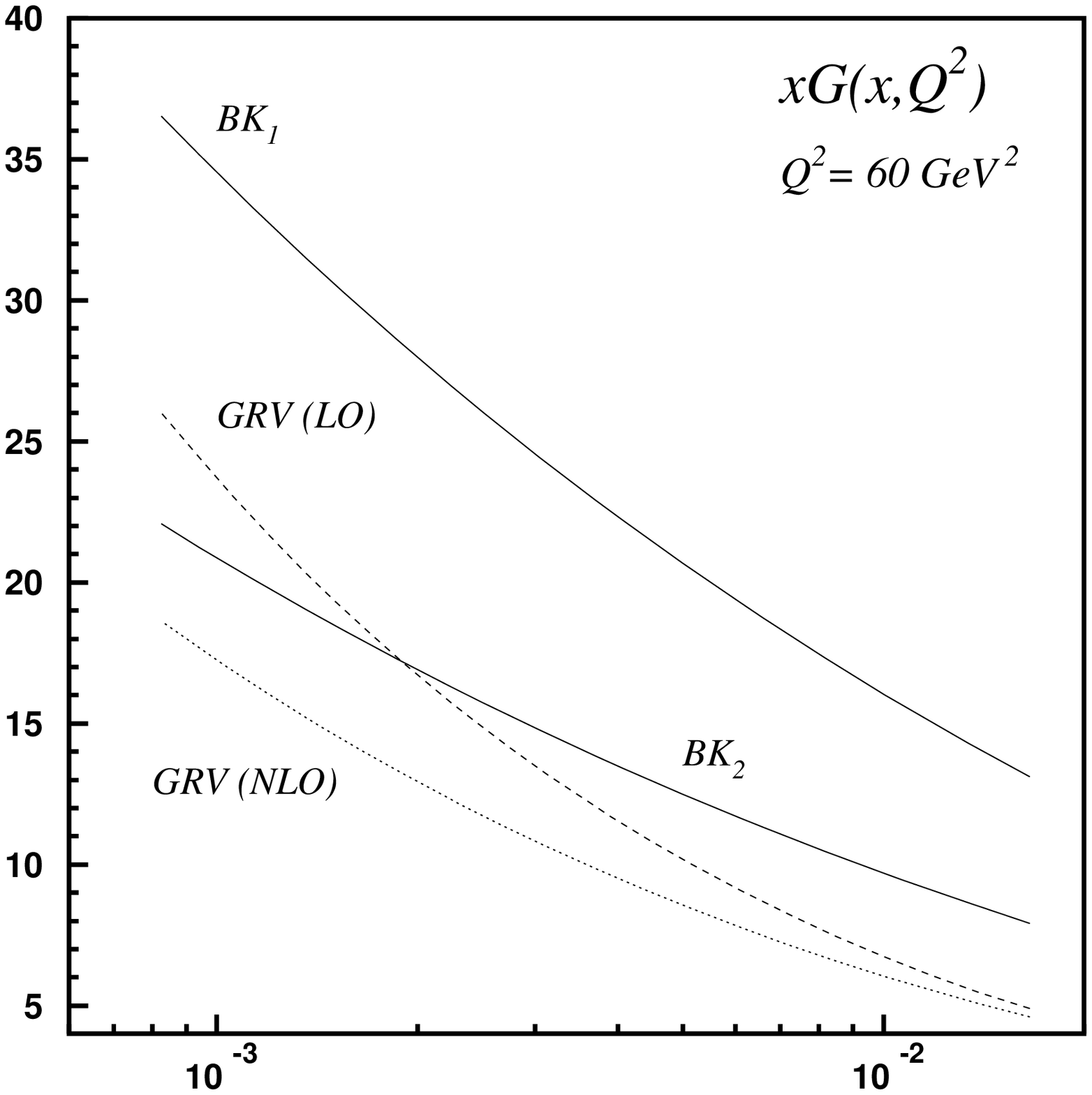,width=85mm} & \,
\psfig{file= 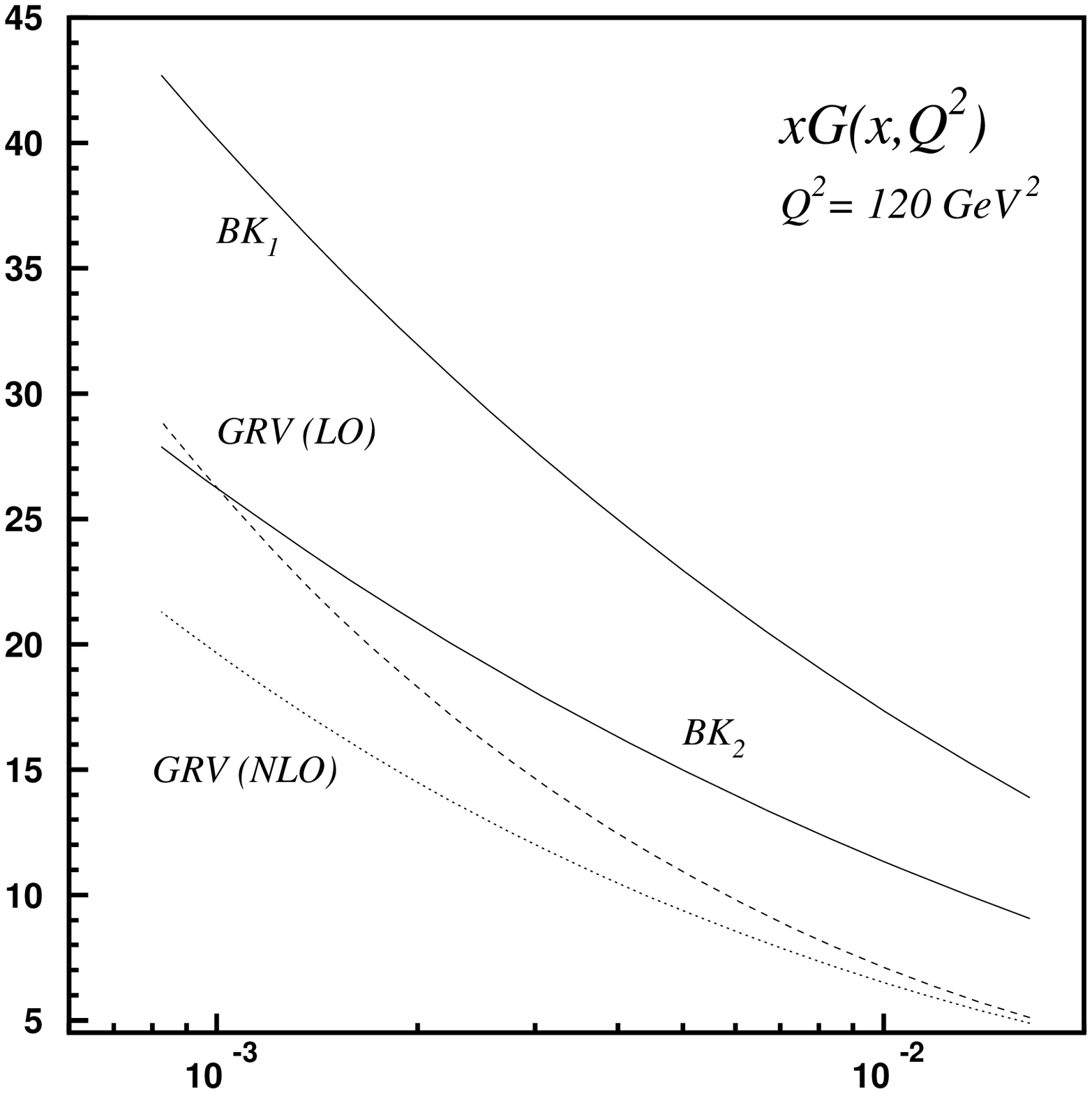,width=85mm}\\
 &  \\ 
\end{tabular}
\caption{\it The integrated gluon density function:
the present model results (solid lines),
LO GRV parameterization (dashed lines) and NLO GRV (doted lines)
as functions of $x$.}
\label{Ungluon}
\end{figure}

 In order to estimate the possible effects of the
variables change Eq.\ref{F4} it is instructive to calculate the
values of integrated gluon density. Indeed, as it seems from
Fig~\ref{F2data} the LO and NLO GRV parameterization give very
close results for the $F_2$ structure function. In the same time,
the integrated gluon density functions are very different
for LO and NLO GRV curves at the same values of $Q^2$.
We present the obtained plots in
Fig~\ref{Ungluon} for the  integrated gluon density functions at
different values of $Q^2$. 

 There are two possible resulting plots of the model in Fig~\ref{Ungluon},
which we denotes as $BK_1$ and $BK_2$ curves. The reason for a existing
of two curves is a following. Let's consider the definition
of integrated gluon density function
\beq\label{F11}
xG(x,Q^2)\,=\,\int\,d^2\,b\,\int^{Q^2}\,\frac{d\,k^2}{k^2}\,
f(x,k^2,b)
\eeq
Due the variable change Eq.\ref{F4} we obtain
a integrated gluon density function in terms of the new
function $\tilde{f}$
\beq\label{F12}
xG(x,Q^2)\,=\,\int\,d^2\,b\,\int^{Q^2}\,\frac{d\,k^2}{k^2}\,
\frac{\tilde{f}(x,k^2,b)}{\alpha_s(k^2)}\,\rightarrow\,
\frac{1}{\alpha_s}
\int\,d^2\,b\,\int^{Q^2}\,\frac{d\,k^2}{k^2}\,
\tilde{f}(x,k^2,b) \,
\eeq
From the Eq.\ref{F12} it is clear, that the
definition of the $xG(x,Q^2)$ in terms of $\tilde{f}$
has a ambiguities in LO scheme calculations due the running
coupling $\alpha_s(k^2)$ under the integral over $k^2$. In our
LO calculations we need to choose the fixed LO value of the $\alpha_s(k^2)$
and to extract it from the integration over $k^2$ in Eq.\ref{F12}.
Therefore, we considered  two possibilities for the
fixed $\alpha_s$ value. As the first one we took the value of $\alpha_s$
from the Table~\ref{Param}, obtained in the fitting of the data.
This choice results are denoted as $BK_1$ curves in the Fig~\ref{Ungluon}.
Another choice
for the $\alpha_s$ is the values of  $\alpha_s$ from the
NLO GRV parameterization taken separately for coresponding $Q^2$.
The results for this value of $\alpha_s$ is  denoted as $BK_2$  in the Fig~\ref{Ungluon}.

 The simple conclusions, which could be immediately
obtained from the present plots, are the following. First of all,
the NLO GRV value of $\alpha_s$ for each $Q^2$ in Eq.\ref{F12}
gives more correct value of $xG(x,Q^2)$ comparing with the common
$\alpha_s$ value from the data fit.
This fact related with the use of the LO $\tilde{f}$ function in our scheme
of calculations
and, therefore, each calculation of the $xG(x,Q^2)$ needs the redefinition
of the present $\alpha_s$ value. The second conclusion concerns the
shape of the found curves. It is easy to see, that both $BK_1$
and $BK_2$ curves have a shapes similar to the NLO GRV curve and pretty different
from the LO GRV curve for integrated gluon density.
This is a sign, that
the simple redefinition of the variables Eq.\ref{F4} in BK equation
allows to include a some part of NLO corrections
to the integrated gluon density and $F_2$ functions. It is
important to underline again, that obtained integrated gluon density function
is similar to the integrated gluon density function obtained with the use of
GRV parameterization.

\section{Conclusion}

 We demonstrated, that based on QCD BFKL pomerons BK evolution equation
for unintegrated gluon density with impact parameter dependence
could be used as a calculation tool for the $F_2$ structure function
and integrated gluon density function. In the large range of energies and
large range of values of $Q^2$ we obtained a good description of DIS data
for $F_2$ structure function. We note, that the obtained
results are in good agreement with results obtained with the help of GRV
parameterization of parton densities and therefore could be used as
independent parameterization of unintegrated gluon density.
It is important, because for our framework it means
that we not only reproduced the results for DIS using more
complex theory than usual evolution equations without impact
parameter dependence, but also that we found the initial conditions for the
proton-proton scattering in the framework of Braun equations
\cite{braun1,braun2,bom}. Therefore, 
the obtained impact parameter dependent parameterization of proton shape
Eq.\ref{F10} with parameters of Table~\ref{Param} allow to
apply formalism  of \cite{bom1} to the important and more general case
of proton-proton scattering, see \cite{bom3}. Another interesting field of the 
application of the proposed model, is the description of the processes of exclusive 
particle production. As it was shown in \cite{bom4}, 
the account of impact parameter 
dependence of the proton-proton scattering amplitude is very important
for the better understanding and better description of the  low  momentum region 
and NLL corrections in the resulting amplitude of the process of exclusive Higgs production.

 The unexpected result, obtained in present calculations, it is a bad agreement
between the calculations in our approach and results of similar approaches
in description of $F_2$ function at small $Q^2\,=\,0.25\,GeV^2$.
Usually, this region of the small values of $\,Q^2\,$
at high energies is considered as a region where the saturation effects are large,
see \cite{GolecBiernat1}. In our case, as it seems, the evolution over rapidity
in impact parameter space does not lead to the saturation effects
which will generate appropriate slope of $F_2$ function at  very small $Q^2$ and
small $x$. The reasons for such a distinction from usual saturation models behavior
is not clear. The including of NLO corrections into the
calculation scheme could, in principal, to improve the situation. At small
values of $Q^2$ the effect of NLO corrections must be large, it is clear if
we will consider the averaged value of $\alpha_s$ obtained in our fit.
This value is very small
and in the theory with running coupling constant at small $Q^2$
this value must be changed a lot, giving a more appropriate result for $F_2$ at small $Q^2$.

 Another approach to this problem, is that
the DIS process at small $Q^2$
physically is very similar to the hadron-hadron scattering,
see \cite{Bartels2} for example.
From this point of view it is not clear why the simple "fan" structure of
BK equation must work well at small values of $Q^2$.
More complicated
"net" diagrams of interacting pomerons became to be important
in this case , see
\cite{braun1,braun2,bom1,Bond,Bond1}, and
absence of these diagrams in BK equation could lead to the
wrong results for DIS at small $Q^2$. Interesting to note, that from the formal point
of view these "net" diagrams are also part of NLO correction to the unintegrated gluon density, which arise from the field theory part of the process and not from the  corrections to the BFKL kernel. We plan to investigate this question in our future studies of the gluon density function in the framework with NLO corrections included.


\section*{Acknowledgments}

\,\,\,I am  especially grateful to Y.Shabelsky for the  discussion
on the subject of the paper and to Leszek Motyka for the
help and useful comments.
This work was done with the support of the Ministerio de Educacion
y Ciencia of Spain
under project FPA2005-01963 together with Xunta de Galicia
(Conselleria de Educacion).



\begin{thebibliography}{100}


%
\bibitem{bfkl}
L.~N.~Lipatov, {Phys.\ Rept.\ } {\bf  286}  131(1997) ;
Phys.\ Rept.\  {\bf 100}  1 (1983) .
%


%
\bibitem{braun1}  M.~A.~Braun,  Phys.\ Lett.\ B {\bf 483}  115 (2000).
%
\bibitem{braun2}  M.~A.~Braun,  Eur.\ Phys.\ J.\ C {\bf 33}  113 (2004) .
%
\bibitem{bom4}
  J.~Bartels, S.~Bondarenko, K.~Kutak and L.~Motyka,
  Phys.\ Rev.\  D {\bf 73}, 093004 (2006).
%
\bibitem{bom}  S.~Bondarenko and L.~Motyka,  Phys.\ Rev.\  D {\bf 75},
 114015 (2007) .
%
\bibitem{pryl}
  E.~Levin and A.~Prygarin,
  arXiv:hep-ph/0701178.
%
\bibitem{bom1}
 S.~Bondarenko, Nucl. Phys. A {\bf 792} 264 (2007).
%
\bibitem{Bond}
  S.~Bondarenko and M.~A.~Braun,
  Nucl.\ Phys.\  A {\bf 799}, 151 (2008)
%
\bibitem{bom2}
S.~Bondarenko and A.~Prygarin Nucl. Phys. A {\bf 800}  63 (2008).

%
\bibitem{bal} I.I.Balitsky, Nucl. Phys. {\bf B463} 99 (1996) .
%
\bibitem{kov} Yu.V.Kovchegov, Phys. Rev. {\bf D60}  034008 (1999) ;
{\bf D61} 074018 (2000) .
%


\bibitem{GolecBiernat1}
  K.~J.~Golec-Biernat and M.~Wusthoff,
  Phys.\ Rev.\  D {\bf 59}, 014017 (1999)

\bibitem{GolecBiernat2}
  K.~J.~Golec-Biernat and M.~Wusthoff,
  Phys.\ Rev.\  D {\bf 60}, 114023 (1999).
\bibitem{GolecBiernat3}
  J.~Bartels, K.~J.~Golec-Biernat and H.~Kowalski,
  Phys.\ Rev.\  D {\bf 66}, 014001 (2002).
%
\bibitem{Mueller1}
  A.~H.~Mueller,
  Nucl.\ Phys.\  B {\bf 335}, 115 (1990).
%
\bibitem{Munier1}
  S.~Munier, A.~M.~Stasto and A.~H.~Mueller,
  Nucl.\ Phys.\  B {\bf 603}, 427 (2001).
%
\bibitem{Lev1}
  E.~Levin and M.~Lublinsky,
  Phys.\ Lett.\  B {\bf 521}, 233 (2001).


\bibitem{Lev2}
  E.~Gotsman, E.~Levin, M.~Lublinsky and U.~Maor,
  Eur.\ Phys.\ J.\  C {\bf 27}, 411 (2003).

\bibitem{Lev3}
  E.~Levin and M.~Lublinsky,
  Nucl.\ Phys.\  A {\bf 712}, 95 (2002).

\bibitem{Bartels1}
  J.~Bartels, E.~Gotsman, E.~Levin, M.~Lublinsky and U.~Maor,
  Phys.\ Rev.\  D {\bf 68}, 054008 (2003).

\bibitem{Kowalski1}
  H.~Kowalski and D.~Teaney,
  Phys.\ Rev.\  D {\bf 68}, 114005 (2003).

\bibitem{Iancu1}
  E.~Iancu, K.~Itakura and S.~Munier,
  Phys.\ Lett.\  B {\bf 590}, 199 (2004).

\bibitem{Kowalski2}
  H.~Kowalski, L.~Motyka and G.~Watt,
  Phys.\ Rev.\  D {\bf 74}, 074016 (2006).

\bibitem{Kut1}
  K.~Kutak and J.~Kwiecinski,
  Eur.\ Phys.\ J.\  C {\bf 29}, 521 (2003).

\bibitem{Kut2}
  K.~Kutak and A.~M.~Stasto,
  Eur.\ Phys.\ J.\  C {\bf 41}, 343 (2005).


\bibitem{Kut3}
  K.~Kutak,
  arXiv:hep-ph/0703068.


\bibitem{GRV}
  M.~Gluck, E.~Reya and A.~Vogt,
  Eur.\ Phys.\ J.\  C {\bf 5}, 461 (1998).

\bibitem{Kwi1}
  J.~Kwiecinski, A.~D.~Martin and A.~M.~Stasto,
  arXiv:hep-ph/9706455.


\bibitem{DIS}
  J.~Breitweg {\it et al.}  [ZEUS Collaboration],
  Phys.\ Lett.\  B {\bf 487}, 53 (2000);
 C.Adloff{\it et al.}  [H1 Collab],
 Eur.\ Phys.\ J.\  C {\bf 21}, 3 (2001);
  S.~Chekanov {\it et al.}
  [ZEUS Collaboration],
  Eur.\ Phys.\ J.\  C {\bf 21}, 443 (2001);
  S.~Chekanov {\it et al.}
  [ZEUS Collaboration],
  Nucl.\ Phys.\ J.\  B {\bf 695}, 3 (2004).


\bibitem{STEQ}
H.L. Lai {\it et al.}, Phys. Rev. D{\bf 55}, 1280 (1997).


\bibitem{Bartels2}
  J.~Bartels and H.~Kowalski,
  Eur.\ Phys.\ J.\  C {\bf 19}, 693 (2001).
%
\bibitem{Bond1}
  S.~Bondarenko, E.~Gotsman, E.~Levin and U.~Maor,
  Nucl.\ Phys.\  A {\bf 683}, 649 (2001).
%
\bibitem{bom3}  S.~Bondarenko and L.~Motyka, in preparation.
%

\end{thebibliography}
\end{document}